 \documentclass{emulateapj}

\def\h2o{H$_2$O}

\def\ch4{CH$_4$}

\def\arcs{\ifmmode {''}\else $''$\fi}

\received{2007 October 8}
\slugcomment{Accepted to ApJ}
\begin{document}

\title{A SURVEY OF 3.3 $\mu$m PAH EMISSION IN PLANETARY NEBULAE}

\author{ERIN C. D. SMITH\altaffilmark{1}, IAN S. MCLEAN\altaffilmark{2}}

\altaffiltext{1}{Department of Physics and Astronomy, UCLA, Los Angeles, CA 90095-1562; erincds@astro.ucla.edu}

\altaffiltext{2}{Department of Physics and Astronomy, UCLA, Los Angeles, CA 90095-1562; mclean@astro.ucla.edu}

\begin{abstract}

Results are presented from a pilot survey of 3.3 $\mu$m PAH
emission from planetary nebulae using FLITECAM, an instrument
intended for airborne astronomy with SOFIA. The observations were
made during ground-based commissioning of FLITECAM's spectroscopic
mode at the 3-m Shane telescope at Lick Observatory. Direct-ruled
KRS-5 grisms were used to give a resolving power (R)$\sim$1,700. 
Targets were selected from IRAS,
KAO and ISO sources with previously observed PAH emission at
longer wavelengths. AGB stars and PN with C$/$O ratios $<$ 1 were
also added to the target list in order to test PAH detection
thresholds. In all, 20 objects were observed.  PAH emission was
detected in 11 out of 20 observed targets.

\end{abstract}


\section{Introduction}

The broad infrared emission features observed at 3.3, 6.2, 7.7,
8.6, and 11.2 $\mu$m have been the subject of astronomical
investigation since their first observation in the 1970s (Gillett et al. 1973; Gillett
et al. 1975). Through comparison to laboratory spectra and
additional observational data these bands have been attributed to
bending and stretching modes of polycyclic aromatic hydrocarbons
(PAHs) (Sellgren 1984; L\'{e}ger \& Puget 1984; Geballe et al. 1985). PAHs are formed
by aromatic structures of carbon and hydrogen which can bind
together into larger assemblies, forming a family of small
molecules such as pyrene, coronene and others.  Emission features
associated with PAHs have been observed in a wide variety of
objects, including planetary nebulae, HII regions, proto-stellar
clouds and star-forming galaxies (Peeters et al.  2002; Hony 2002). This
ubiquity shows PAHs to be potentially important components of the
ISM, and has led to several studies aimed at correlating PAH
emission with region characteristics (Tokunaga et al. 1991; Roche
et al. 1996; Cohen \& Barlow 2005; Peeters et al. 2002; van
Diedenhoven et al. 2004). Recent surveys (Peeters et al. 2002, van Diedenhoven et al. 2004) using space-based telescopes have
shown variability in the PAH emission. Peeters et al. (2002)
introduced a classification scheme which grouped  6.2, 7.7 and 8.6
$\mu$m PAH spectra into three groups based on band FWHM, peak
wavelength and overall shape. HII regions, reflection nebulae, and
young stellar objects tended to exhibit Class A$_{6-9}$ PAH behavior,
while evolved stellar objects like planetary nebulae and post-AGB
stars exhibited class B$_{6-9}$ emission. A small number of objects,
tending to be very young AGB stars, exhibited class C$_{6-9}$ PAH
emission (Sloan et al. 2007). Van Diedenhoven et al. (2004) expanded this
classification to the 3.3$\mu$m and 11.2$\mu$m bands. Variations
in the 3.3 micron feature were found to be smaller than those in
the 7.7 and 6.2 micron features. This is explained by the fact
that the 3.3 and 11.2 $\mu$m features are due to stretching and
out-of-plane bending modes of the C-H bonds in the PAH molecules,
while the 6.2 and 7.7 $\mu$m features arise from stretching in the
C-C bonds (Allamandola et al. 1989).  Tokunaga et al. (1991) and
van Diedenhoven et al. (2004) classified the 3.3 micron feature
into two classes: class A$_{3.3}$, or class 1, with emission centered at 3.290$\mu$m
and a FWHM of 0.040$\mu$m, and class B$_{3.3}$, or class 2, which have a FWHM of
0.037$\mu$m with emission centered at 3.293$\mu$m (class B1$_{3.3}$) or
3.297$\mu$m (class B2$_{3.3}$). Planetary nebulae and their evolutionary
precursors tended to be class A$_{3.3}$ objects. While other emission
features are sensitive to PAH charge state, comparisons with
laboratory data show the 3.3$\mu$m feature to arise from neutral
PAHs, rather than PAH cations (Allamandola et al. 1989).

PAHs are produced in the hot outflows accompanying the last stages
of evolution of low and intermediate-mass stars. As a star evolves
off the Asymptotic Giant Branch (AGB), a period of dramatic mass
loss begins. Roche et al. (1996) found a threshold C$/$O ratio for
3.3$\mu$m PAH emission in planetary nebulae of $\sim$0.6. A
study of 7 and 8$\mu$m emission by Cohen \& Barlow (2005) found a
roughly consistent threshold ratio of 0.56$\pm$0.3. In nebulae
with lower C$/$O ratios, nebular carbon is expected to be
incorporated in silicates and CO, while carbon-rich nebulae
produce hydrocarbons including PAHs (Allamandola et al. 1989). PAH
emission is caused by UV excitation. In planetary nebulae the
stellar remnant bathes the expanding gas shell with UV photons
which produce elemental nebular lines and also excite the
vibrational and stretching modes of PAH molecules. The relaxation
of these modes produces the broad infrared emission features which
have been observed in several planetary nebulae, as well as their
evolutionary precursors: proto-planetary nebulae and post-AGB
stars (Matsuura et al. 2005; Justtanont  et al. 1996).

This survey has three purposes: investigate the properties of 3.3
$\mu$m PAH emission in planetary nebulae by expanding the analysis
performed by Roche et al. (1996) and Cohen \& Barlow (2005); expand the
classification of 3.3 micron emission to a larger sample of
planetary nebulae using a higher resolution than previous surveys;
and demonstrate the FLITECAM instrument's spectroscopy mode in a
survey likely to be undertaken once aboard SOFIA.

As part of the science verification of the FLITECAM instrument for
SOFIA, we proposed spectroscopic and narrow band imaging
observations of the 3.3 $\mu$m PAH feature, together with atomic
and molecular emission, in post-AGB stars, planetary nebulae, HII
regions and reflection nebulae. With the significantly increased
sensitivity of stratospheric observations, the overall goal of
this project is to trace the origin and evolution of PAHs from
their birth environment through their processing in the ISM to
their incorporation in proto-stellar material. The 3.3$\mu$m
feature is quite broad and typically resolved at R$\sim$300.
FLITECAM's grism mode has a higher spectral resolution that allows
investigation of not only the PAH feature itself, but also PAH
interaction with ionized and neutral gas. In principle, higher
resolution allows subtle changes in PAH emission shape to be
observed. In addition, FLITECAM's narrow-band imaging capability
allows one to study the spatial distribution of PAH emission in
spatially-resolved objects. FLITECAM offers three advantages for
this study: a wide field of view ($\sim$8 arc minutes diameter)
for imaging studies, good spectral resolution (R$\sim$1700) and,
when used with SOFIA, a reduced water vapor overburden and colder
telescope thus significantly reducing the background at this
wavelength.

Prior to the start of SOFIA flight operations we decided to try a
pilot-study from a ground-based observatory in order to develop
the methodology for this project. Although the low water-vapor
advantage is completely lost and backgrounds are much increased
for a ground-based telescope, we have successfully detected the
3.3 $\mu$m PAH feature, with relatively high spectral resolution,
in several planetary and proto-planetary nebulae. Wherever
possible, the strength of the emission has been quantified and
compared with other physical properties such as C$/$O ratio. The
pilot study demonstrates the potential for more sensitive PAH
studies when SOFIA is operational. Our results and methods,
including a brief description of the instrument, are described in
this paper.

Target selection, the FLITECAM grism mode and the observations are
described in \S2, data reduction methods are explained in \S3 and
results are presented in \S4. Analysis and conclusions are given
in \S5.

\section{Observations}

\subsection{Target Selection}

\begin{deluxetable*}{lccccc}
\tablecaption{\bf TARGET LIST AND OBSERVING LOG \label{tbl-1}} 
\tabletypesize{\scriptsize}
\tablenum{1}
\tablehead{ \colhead{$ $} &\colhead{R.A.} &\colhead{Dec.} &\colhead{Object} &\colhead{PAH} &\colhead{UT Date}\\
\colhead{Object} &\colhead{(J2000.0)} &\colhead{(J2000.0)} &\colhead{Type} &\colhead{Detected?\tablenotemark{(a)}} &\colhead{of Observation}}
\startdata

M 3$-$35            &20 21 03.769       &$+$32 29 23.86     &PN       &Y    &July 25 2005\\
NGC 7027            &21 07 01.593        &$+$42 14 10.18    &PN        &Y    & Oct 23 2005\\
HB 5                &17 47 56.187        &$-$29 59 41.91    &PN        &Y     &May 9 2006\\
IC 5117             &21 32 31.027        &$+$44 35 48.53    &PN        &Y     &July 22 2005\\
BD+30$^{\rm o}$3639        &19 34 45.2323      &$+$30 30 58.936    &PN        &Y    &May 6 2004\\
M 2$-$43            &18 26 40.048       &$-$02 42 57.63     &PN        &Y    &July 25 2005\\
IRAS21282$+$5050\tablenotemark{(b)}    &21 29 58.42        &$+$51 03 59.8      &PN        &Y     &July 23 2005\\
Hen 2$-$459         &20 13 57.898       &$+$29 33 55.94     &PN        &Y    &July 23 2005\\
WHME$-$1            &19 14 59.755       &$+$17 22 46.01     &PN        &$?$     &July 22 2005\\
J900\tablenotemark{(c)}   &06 25 57.275       &$+$17 47 27.19     &PN        &$?$     &Oct 8 2006\\
NGC 6886            &20 12 42.813       &$+$19 59 22.65     &PN        &Y     &Oct 8 2006\\
NGC 6790            &19 22 56.965       &$+$01 30 46.45     &PN        &Y    &May 7 2006\\
M 4$-$18            &04 25 50.831       &$+$60 07 12.72    &PN        &Y    &Oct 22 2005\\
AFGL 3068     &23 19 12.39 &$+$17 11 35.4    &Mira        &Cont    &Oct 20 2005\\
AFGL 3116    &23 34 27.66       &$+$43 33 02.4   &C*       &Cont    &Oct 21 2005\\
AFGL 337    &02 29 15.3       &$-$26 05 55.67    &Mira        &Cont    &Oct 22 2005\\
YY Tri     &02 18 06.6       &$+$28 36 48.3    &C*        &Cont   & Oct 07 2006\\
M 2-9     & 17 05 37.952 & $-$10 08 34.58  & PN (O-rich) &N& June 4 2004\\
NGC 3242 & 10 24 46.107   &$-$18 38 32.64 &  PN (O-rich) & N& May 9 2006\\
NGC 6210 & 16 44 29.491   &$+$23 47 59.68 & PN (O-rich) & N & May 7 2006\\

\enddata

\tablenotetext {a}{Cont: Continuum Object; C*: Carbon-rich AGB star; o-rich: oxygen-rich chemistry; ?: possible PAH detection Y: PAH detected, N: No PAH detection}
\tablenotetext {b}{IRAS21282+5050 is also named PN G093.9$-$00.1}
\tablenotetext{c}{J900 is also PN G 194.2$+$02.5}

\end{deluxetable*}
Targets for our 3.3$\mu$m survey were selected from ISO, KAO and
IRAS observations of planetary nebulae, proto-planetary nebulae
and post-AGB stars. Our sample was compiled based on the reported
presence of PAH emission in longer wavelength bands observed from
space, as well as practical issues of signal-to-noise limitations
at a low-altitude ground-based site. Of the 20 objects in the
survey, 13 had previous long-wave PAH detection in at least one
band (Cohen et al. 1989; Jourdain de Muizon et al. 1990; Rinehart et al. 2002). Additional sources were added in order to extend the sample
to post-AGB stars and proto-planetary nebulae. Since PAH emission
is expected only in objects with a carbon-rich dust chemistry, all
but 3 objects have C$/$O ratios above the threshold ratio for PAH emission determined by Roche et al. and Cohen \& Barlow. One O$-$rich nebula shows a somewhat
mixed dust chemistry, while the remaining two O$-$rich objects
were used as a test of methodology (Jourdain de Muizon et al. 1990). The targets in this sample
represent a somewhat larger range in morphology, size and
brightness than would normally be selected for a survey of this
type. This was done in order to meet the broad goals of the pilot
study and to facilitate the fullest possible science verification
of the FLITECAM grism mode.

All data presented here were obtained using the Shane 3-meter
telescope at Lick Observatory, Mt. Hamilton, CA, with the FLITECAM
instrument attached. Observations were collected over the course
of five separate observing runs from May 2004 to October 2006.
Table 1 lists all the targets, total observation times, and the
standard stars used to remove telluric absorption lines. Shorthand
names (such as BD30 for BD+30$^{\rm o}$3639) are used throughout the
text for simplicity, but full names are given in Table 1.

\subsection{Instrumentation}

As mentioned above, our pilot-study of 3.3$\mu$m emission from PN
was carried out as part of the commissioning and performance
verification of a new instrument called FLITECAM (First Light Test
Experiment Camera). FLITECAM is a 1-5$\mu$m camera developed by
us at the UCLA Infrared Laboratory (P.I.: McLean) for NASA's SOFIA
(Stratospheric Observatory for Infrared Astronomy) project. SOFIA
is a modified Boeing 747-SP airplane with a 2.5-meter f$/$19.6
bent-Cassegrain telescope operating at altitudes up to 45,000 ft
and therefore above 99\% of the atmosphere's water vapor content.
Most of the SOFIA instruments are designed for much longer
wavelengths, but because FLITECAM operates in the near-infrared,
and because the f$/$17 optics of the Lick 3-meter telescope
provides a plate scale similar to that of SOFIA, FLITECAM can be
used from the ground without additional optics. Performance and
technical details of the instrument are described in McLean et al.
2006. First light was obtained in October 2002 at Lick Observatory
and the instrument has accumulated data with the 3-meter telescope
on eight separate occasions. Photometric results from a narrow
band survey of nearby star-forming regions for methane-bearing
planetary mass objects have been published elsewhere (Mainzer and
McLean 2003; Mainzer et al. 2004; McLean et al. 2006).

Briefly, FLITECAM utilizes large refractive optics to inscribe an
$\sim$8 arc minute diameter FOV on a 1024x1024 InSb ALADDIN III
detector with a plate scale of just under 0.5 arc seconds per
pixel. The instrument is cooled with 20 liters of LN$_{2}$ and 20
liters of LHe and can be operated in either the up-looking
position (Lick) or the horizontal position (SOFIA). Another
feature is a dual filter wheel which houses standard broadband
filters, several narrow band filters plus a suite of three
direct-ruled KRS-5 grisms and their appropriate order-sorting
filters. Each grism can be used in one of three orders giving a
total of 9 settings to cover the entire 1-5 $\mu$m region. The
KRS-5 grisms provide a moderate resolution spectroscopy mode in
conjunction with fixed slits of either 1\arcs~ or 2\arcs~ width
and 60\arcs~ total length to yield resolving powers of R$\sim$1700
and 900 respectively (Smith and McLean 2006). When airborne,
FLITECAM will be the only instrument providing spectral coverage
from 2.5-5.5$\mu$m with a low water vapor and low background
environment.


\subsection{Observations}
FLITECAM's imaging mode was used to place each object on the
selected slit, and the observatory's offset CCD guide camera was
used to hold the source on the slit. Spectroscopic observations
were made in a standard ABBA nod pattern. Typically, integrations
of 300 s each (3 s exposure with 100 coadds) were taken with the
object placed at two positions, designated A and B, separated by
$\sim$26$\arcs$ on the $\sim$60$\arcs$ long entrance slit of
FLITECAM. Shorter exposure times were used for brighter objects.
The total number of ABBA cycles depends on the object, but all
targets were observed for a minimum of one cycle. Total
integration times per object ranged from 20 minutes to 1.5 hours.
Signal-to-noise ratios were typically greater than 10 (10\%) per
resolution element. Seeing conditions were $\sim1\farcs0-1\farcs2$
and a slit width of 1$\farcs$0 ($\sim$2 pixels) was used for all
observations, except in the case of HB~5 for which we used a
$2\farcs0$ (4 pixels) slit because of poorer seeing.

The grism mode selected provides coverage from 3.0-3.4$\mu$m. This
spectral region exhibits significant contamination from absorption
lines of terrestrial water vapor, making ground-based observations
very difficult. In principle, these absorption features can be
removed by observations of standard stars. Two methods were used
to obtain satisfactory telluric corrections; 1) ATRAN models of
the atmosphere (Lord 1979); 2) observations of a bright A0V star
or MIII giant star at an airmass very close to that of the target
object. Great care was taken to ensure the telluric standard did
not saturate the FLITECAM detector. Arc lamp spectra, taken at the
end of the night, and OH night-sky lines in the observed spectra,
were used for wavelength and dispersion calibration. A white-light
spectrum and a corresponding dark frame were also obtained for
flat-fielding. Calibration stars are listed in Table 1.

\section{Data Reduction Methods}
\subsection{Redspec}
Survey data were reduced using REDSPEC, an IDL-based software
package developed at UCLA for spectra obtained with the Keck
NIRSPEC instrument by S. Kim, L. Prato and I. McLean\footnote{See
http://www2.keck.hawaii.edu/inst/nirspec/redspec/index.html}.
REDSPEC was modified for use on FLITECAM data by adding routines
to flip the raw spectra, to automatically subtract nod pairs to
enable spatial rectification of thermal-IR spectra (in which the
traces are hard to see against the background), and to apply
FLITECAM wavelength solutions to the spectra.

To extract spectra free from atmospheric background and uneven
detector response, the difference of an A/B image pair was formed
and flat-fielded. The flat-fielded difference frame was then
rectified using the spatial and spectral maps and the raw spectrum
produced by summing 5$-$10 rows from each trace in the rectified
image. The extracted traces (one positive, one negative) are then
subtracted again to produce a positive spectrum with residual
night-sky emission line features removed, unless a line was
saturated.  Telluric standard star spectra were reduced in the
same way. After interpolating over the intrinsic hydrogen
absorption lines, the reduced target spectrum was divided by the reduced
calibrator star spectrum to remove telluric features. The true
slope of the target spectrum was restored by multiplication with a
blackbody spectrum of T$_{eff}=9500$ K for an A0 V star, and
T$_{eff}=3000$ K for an MIII star(Tokunaga 1991). Finally, the
spectra reduced from multiple A/B pairs were averaged together to
improve the signal-to-noise ratio.

\subsection{ATRAN telluric correction}
Ground-based thermal infrared observations are dramatically
affected by variability in atmospheric water vapor and background
emission. Therefore, good telluric correction is critical to
accurate reduction of thermal IR spectra. In many cases,
appropriate telluric standards are either too faint for good
correction, or too widely separated for a good airmass match. In
addition, the infrared backgrounds at Lick observatory can change
during observations of target objects, making accurate telluric
correction using observed standards very difficult. As an
alternative to observed telluric standards, ATRAN, an atmospheric
modelling program developed by Steve Lord, was used to generate
artificial telluric calibrator spectra at the target's zenith
angle and airmass. ATRAN models allow tweaking of water vapor
over-pressure and other atmospheric conditions to produce
optimized telluric corrections for each target. Spectra of targets
with high S/N ratios (NGC 7027, BD+30$^{\rm o}$3639) showed no appreciable
differences between telluric-corrected spectra using observed
standard stars and ATRAN-corrected spectra. Low S/N object spectra
were noticeably improved using ATRAN-produced telluric standards.
This makes sense because low S/N targets were observed for longer
total time periods than high S/N targets, increasing the
contribution of atmospheric variability in the resulting spectrum.
While optimized ATRAN models can be fit individually to each
target AB pair in a long sequence, the single observed telluric
calibrator is only a good fit for one moment in time.

ATRAN models were computed using a standard 2 layer atmosphere at
4000 ft elevation using the appropriate zenith angle for each
target observation. The water vapor over-pressure was varied from
8 to 12 mm (around the standard predicted value of 10 mm) for each
target object to obtain the best fit to the data. This produced a
normalized, flat-fielded calibrator spectrum. Each target
observation was reduced using the REDSPEC software package as
described in \S3.1, but without division by a calibrator.
Instead, the reduced target spectrum was divided by the
ATRAN-produced telluric calibrator spectrum.

\section{Results}

\subsection{Overview}

Those objects that yielded the best PAH spectra tended to be
young, high-excitation compact planetary nebulae. Objects such as
NGC 3242 for example, were too low in surface brightness to yield
a useful spectrum given the high background of the ground-based
observations. Very bright objects, such as AFGL 3086, also did not
yield a PAH detection because the central object saturated the
FLITECAM detector, even at the shortest possible exposure times,
and any extended PAH emission located far enough away from the
central object so as to avoid saturation was too low in surface
brightness to be detected. All reduced spectra show residual
structure from atmospheric water vapor lines that could not be
removed. L-band observations are difficult at Lick Observatory
because the site is only 4209 ft above sea level, and is located
in a particularly humid area. These conditions lead to highly
variable telluric features, often variable on the timescale of one
nod (5 minutes). While the telluric correction steps described
above removes the majority of these features, even the best
spectrum, that of NGC 7027, shows residual structure. It should be
noted that these complications exist, although to a much reduced
level, even at the world's best astronomical sites, such as Mauna
Kea. However, with the caveat on the limitations of the telluric
corrections, we can analyze the detected PAH emission features.
\begin{figure}
\epsscale{1.15}
\plotone{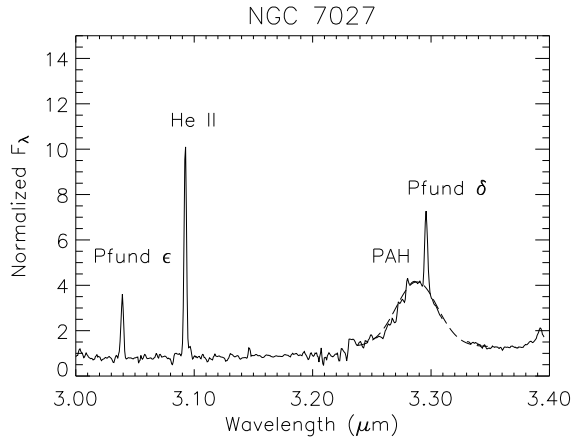}
\caption{Spectrum of NGC 7027 with prominent emission features labelled. The missing data from 3.31 to 3.33 $\mu$m is not plotted due to saturated atmospheric methane and water features. The dashed line shows the gaussian fit to the data.}.\label{fig1}
\end{figure}
Figure 1 shows the extracted 3.3$\mu$m spectrum for NGC 7027. Of
course, NGC 7027 is a particularly bright, high-excitation
planetary nebula with well-known PAH emission in all bands.
Hydrogen and helium lines are seen in emission, as is the broad
PAH feature. One hydrogen line, Pfund delta (3.296$\mu$m) is seen
in emission on top of the broad PAH feature. Also evident is the
effect of residual terrestrial water absorption. The blue slope of
the PAH feature has several unsaturated, but variable water
absorption features, leading to a slightly stair-stepped shape
even on NGC 7027. On the red slope, there is a deep atmospheric
methane line at 3.32$\mu$m. This line cannot be removed from the
spectrum, thus no data points are plotted for the
3.310-3.330$\mu$m region.

The FLITECAM spectra presented here cover from 3.0 to 3.4$\mu$m at
a resolution of R$\sim$1700. After reduction, all spectra
underwent a 5-sigma cut to remove hot pixels and residuals from
saturated atmospheric lines. Figures 2 and 3 show a selection of the
target spectra all plotted in a similar way. There are four
classes of resulting spectra: 1) continuum objects, 2) PAH
non-detections, 3) PAH detections and 4) PAH possible detections.
(See Table 1 again for a list of all observed PNs and their
classification.)

\begin{figure*}
\epsscale{.75}
\plotone{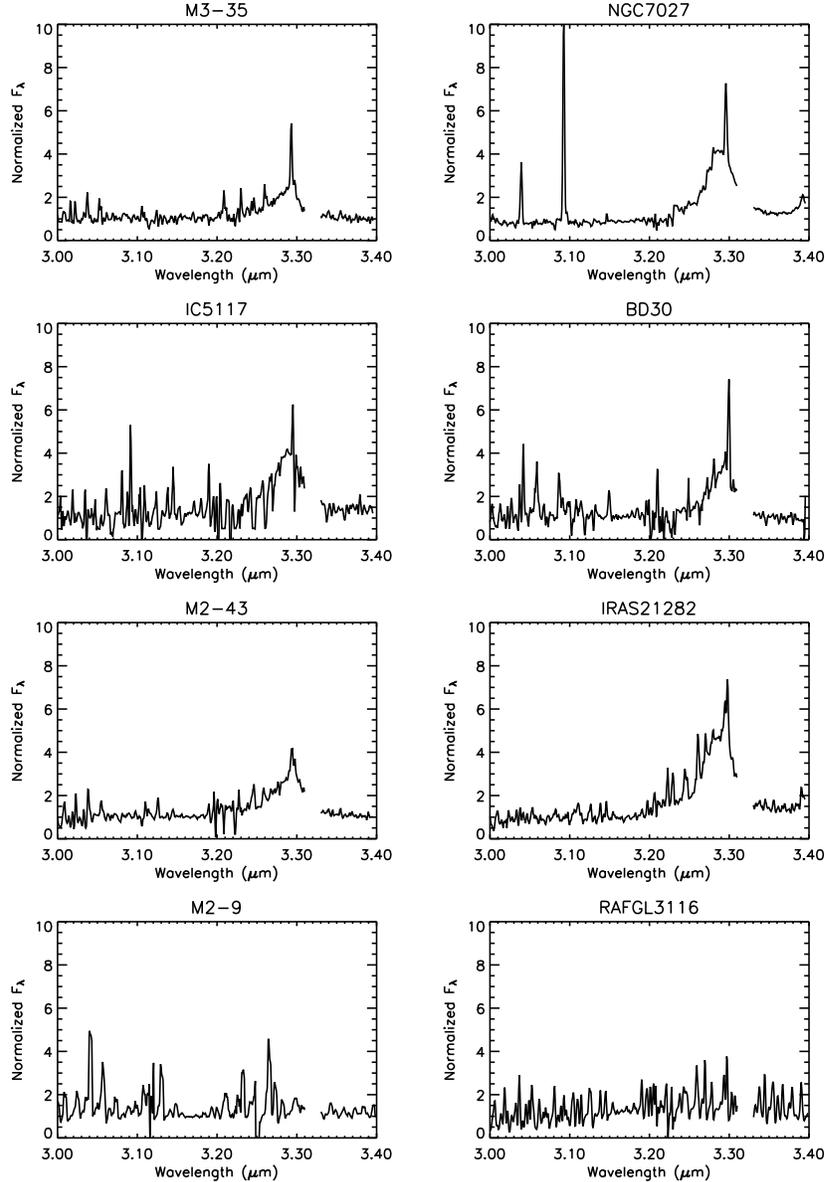}
\caption{Reduced spectra from PAH survey. The missing data from 3.31 to 3.33 $\mu$m is not plotted due to saturated atmospheric methane and water features. Fainter objects show noise spikes in the 3.0 to 3.2 $\mu$m region due to terrestrial absorption lines. Detected emission lines are listed in Table 2. All detected PAH features are well-fitted by a Gaussian curve. IRAS 21282 shows some non-symmetry, but has FWHM and peak PAH wavelength emission consistent with class A$_{3.3}$  emission.}.\label{fig2}
\end{figure*}

\begin{figure*}
\epsscale{.75}
\plotone{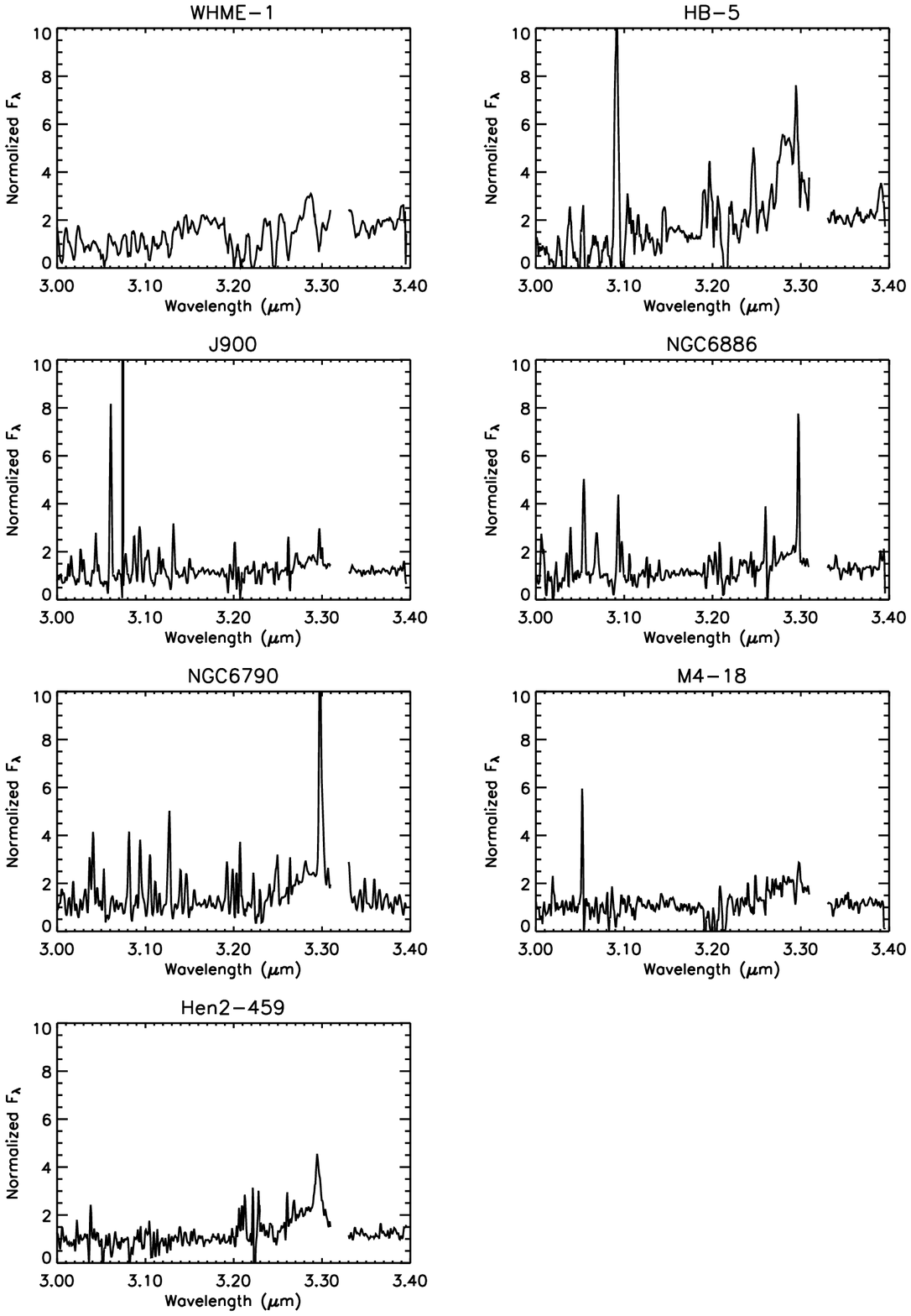}
\caption{Reduced spectra from PAH survey. The missing data from 3.31 to 3.33 $\mu$m is not plotted due to saturated atmospheric methane and water features. Fainter objects show noise spikes in the 3.0 to 3.2 $\mu$m region due to terrestrial absorption lines. Detected emission lines are listed in Table 2. All detected PAH features are well-fitted by a Gaussian curve.}.\label{fig3}
\end{figure*}
\subsection{Continuum objects}
Four members of the sample exhibit a stellar spectrum with no
obvious emission features. All four of these objects are post-AGB
or AGB stars, and are thus less evolved than the rest of the
sample. RAFGL 3116 does have reported PAH detections at longer
wavelengths from ISO observations (Jourdain de Muizon et al. 1990), but does not
show an obvious PAH 3.3$\mu$m band peak in our observations. The
feature may be present, but could be swamped by the strong, red
spectrum of the highly evolved AGB star. The spectrum of RAFGL
3116 in Figure 2 is typical of this type of object.

\subsection{PAH non-detections}
Three members of the sample (M 2-9, NGC 3234, and NGC 6210) exhibit standard PN features (i.e.
atomic emission lines and non-existent or very faint continuum),
but yet show no PAH emission at 3.3$\mu$m.All
three non-detections have an oxygen-rich chemistry, with
C$/$O$<<$1. In these objects, crystalline silicate features tend to dominate, rather than PAH emission (Allamandola 1989). M 2$-$9 in Figure 2 is an example of a oxygen-rich
planetary nebula with no PAH emission.

\subsection{PAH detections and possible detections}
Eleven members of the sample show distinct PAH emission at
3.3$\mu$m, while two more targets (WHME-1 and J900) show
possible PAH emission. PAH flux was quantified by fitting a
gaussian to the PAH band assuming a flat continuum. NGC 7027 is
the best example of a strong PAH detection, and can be seen either
in Figure 1 or Figure 2. WHME-1 and J900 are examples of
possible-detection of PAH emission. While PAH emission is seen at
longer wavelengths in these objects, the nebulae may have been too
faint compared to the high backgrounds at Lick for solid 3.3$\mu$m
detections. Figures 2 and 3 show the extracted spectra for all the PAH
detections. Hydrogen emission at 3.296$\mu$m is evident in many of
the sources with a PAH detection (HB~5, M3-35, BD+30$^{\rm o}$3639 and NGC 7027 in
particular). This profile was subtracted out of the extracted
spectra by fitting a gaussian before fitting a curve to the PAH
band itself in order to determine its strength and shape. Table 2
summarizes the derived PAH measurements for the observed nebulae. Those nebulae with non detections have upper limits on the PAH equivalent widths. The C/O ratios of the nebulae, if known, are also listed.

\section{Discussion and Analysis}
 \begin{deluxetable*}{lccccccc}
\tabletypesize{\scriptsize}
\tablewidth{0pt}
\tablecaption{\bf TARGET PROPERTIES AND PAH RESULTS \label{tbl-2}}
\tablenum{2}
\tablehead{ \colhead{$ $} & \colhead{Class} & \colhead{C/O\tablenotemark{(a)}} & \colhead{PAH EW} & \colhead{PAH FWHM} & \colhead{Cent. Wave.}& \colhead{Other Lines\tablenotemark{(d)}}\\
\colhead{Object} & \colhead{$ $} & \colhead{Ratio} & \colhead{(nm)} & \colhead{($\mu$m)} & \colhead{($\mu$m)}& \colhead{ID (EW(nm))}}

\startdata

M 3$-$35              & A$_{3.3}$        & ---                         & 67.3$\pm$4.9        &0.041$\pm$0.004  & 3.287$\pm$0.002&{\small 3(11) } \\
NGC  7027           & A$_{3.3}$        & 2.15                       & 170.6$\pm$5.7     &0.041$\pm$0.002   & 3.289$\pm$0.002 &{\small 1:(12); 2:(51); 3(32) } \\
HB 5                      & A$_{3.3}$        & ---                       & 80.3$\pm$4.5         &0.042$\pm$0.002  & 3.285$\pm$0.003&{\small 2:(31); 3(9) } \\
IC 5117                 & A$_{3.3}$        & 1.92                     & 113.3$\pm$7.5    &0.038$\pm$0.005 & 3.289$\pm$0.002 &{\small 3(16)} \\
BD+30$^{\rm o}$3639    & A$_{3.3}$          & 1.6                  & 78.6$\pm$8.4     &0.040$\pm$0.003 & 3.289$\pm$0.003 &{\small 3(11)}  \\
M 2$-$43              & A$_{3.3}$                & ---                        & 66.7$\pm$4.9          &0.045$\pm$0.004 & 3.289$\pm$0.002 &{\small 3(11)}  \\
IRAS21282            & A$_{3.3}$                & ---                      & 166.1$\pm$9.8          &0.042$\pm$0.004  & 3.286$\pm$0.002 &{\small 3(15)}  \\
Hen 2-459            & A$_{3.3}$\tablenotemark{(b)}           & ---                        & 46.6$\pm$5.2          &0.041$\pm$0.004 & 3.287$\pm$0.002&{\small 3(26)}   \\
WHME-1               & A$_{3.3}$\tablenotemark{(b)}                & ---                       &14.9$\pm$9.3\tablenotemark{(b)}           &0.062$\pm$0.010  & 3.286$\pm$0.035 &{\small }   \\
J900            &A$_{3.3}$         &4.39                      &14.4$\pm$7.1\tablenotemark{(b)}           &0.041$\pm$0.004 & 3.289$\pm$0.002&{\small 3( $<$5.6) }  \\
NGC 6886            & A$_{3.3}$        &1.3                       & 46.2$\pm$10.8          &0.043$\pm$0.004 & 3.290$\pm$0.003 &{\small 3(27)} \\
NGC 6790            & A$_{3.3}$        &0.82                    & 57.4$\pm$11.4         &0.041$\pm$0.003 & 3.289$\pm$0.002 &{\small 3(37)}  \\
M 4-18                    & A$_{3.3}$         &2.9                    & 60.0$\pm$33.4\tablenotemark{(c)}            &0.047$\pm$0.003  & 3.286$\pm$0.002&{\small }   \\
AFGL 3068                    & ND        &--           & $<$6.2          &---  & ---&{\small }  \\
AFGL 3116                    & ND        &--                    & $<$7.7            &---  & ---   &{\small }  \\
AFGL 337                    & ND         &--                      & $<$7.6         &---  & ---&{\small }  \\
YY Tri                   & ND         &--                     & $<$6.2           &---  & ---   &{\small }  \\
M 2-9                    & ND        &$<$0.5                     &$<$9.2          &---  & ---  &{\small }  \\
NGC 3242                    &ND        &0.67                     & $<$4.3           &---  & ---   &{\small } \\
NGC 6210                    & ND        &0.26                    & $<$9.1           &---  & ---   &{\small }  \\
\enddata
\tablenotetext{a}{C/O ratio references; NGC 7027: Middlemass (1990); IC 5117, NGC 3242:  Kholtygin (1998); NGC 6886: Pwa et al. (1986); NGC 6790, NGC 6210:  Liu et al. (2004); M 4-18: De Marco \& Barlow (2001); M 2-9: Liu et al. (2001)}
\tablenotetext{b}{Borderline PAH detection}
\tablenotetext{c}{Noisy Continuum}
\tablenotetext{d}{1:Pfund $\epsilon$(3.039$\mu$m) , 2:HeII(3.09$\mu$m), 3:Pfund $\delta$(3.29$\mu$m). EW is listed with Line ID.}

\end{deluxetable*}
Prior to further analysis, each detected PAH feature was fitted
with a gaussian to determine the feature's FWHM and central
wavelength. The average central wavelength of the PAH emission
feature was 3.288$\mu$m $\pm$ 0.002$\mu$m, with a FWHM of 0.042
$\mu$m$\pm$ 0.002$\mu$m. This is consistent with surveys of
similar objects that show this feature to have a gaussian profile
centered at 3.289$\mu$m and a FWHM of 0.042$\mu$m (Tokunaga et al.
1991).

For more detailed analysis of the 3.3$\mu$m feature we adopt the
method used by Roche et al. (1996), that is, we quantify PAH
emission strength by determination of the emission band's
equivalent width (EW). This method is superior to absolute flux
measurements as the EW is less affected by atmospheric absorption
and fluctuation. However, the determination of equivalent width
depends on an accurate model of continuum emission for comparison
to the measured feature, and can thus be rendered inaccurate for
objects with weak continua, or in regions where emission plateaus
mask true continuum emission. As this survey was limited in scope
to only the brightest, most compact planetary nebula, most of the
observed objects exhibited continuum emission above the threshold
necessary for good equivalent width determination. The wavelength
regime of this study (3.0-3.4$\mu$m) is also advantageous to
determining continuum emission levels, because the region just
blue of the PAH emission is relatively free of atmospheric
absorption bands, and is constant over a range roughly comparable
to the width of the PAH feature. Three nebulae have unusable
equivalent width measurements due to either weak PAH emission (as
in WHME1 and J900), or noisy continuum spectra (as in M 4-18).

\subsection{Correlating PAH and C/O ratio}

The most well-studied correlation is between the nebular C$/$O
ratio and the PAH emission strength. Current theory predicts that
nebulae with high C$/$O ratios will exhibit PAH emission, while
more oxygen rich nebulae will exhibit silicate features, having
trapped most nebular C$/$O in CO molecules. Figure 4 shows the
derived 3.3$\mu$m emission equivalent width plotted against the
nebular C$/$O ratio obtained from the literature (Kholtygin 1998). Only 7 nebulae
in our sample have known C/O ratios, and only 5 of these nebulae
have usable equivalent widths. PAH emission equivalent widths are
uncertain up to 10 percent, and even the known C$/$O ratios
measurements have large uncertainties.

Figure 4 shows the C$/$O ratio plotted against the five nebulae
with measured equivalent widths. Also plotted is the linear
regression fit to the data. This shows a PAH equivalent width
detection threshold at C$/$O=0.65 $\pm$ 0.28. This result is
within the uncertainties obtained by Roche et. al. (1996), who
performed a similar analysis, and with Cohen \& Barlow (2005), who
measured a C$/$O ratio versus 7.7$\mu$m PAH emission.

\subsection{PAH emission classification}

For further analysis we characterize the PAH emission according to
the system suggested by van Diedenhoven et al. using band shape,
location and FHWM. Class A$_{3.3}$ emission is characterized by symmetric
PAH emission centered at 3.290$\mu$m with a FWHM of 0.040 $\mu$m.
Class B$_{3.3}$ emission is centered at 3.293 or 3.297$\mu$m with a FWHM
of 0.030$\mu$m. In van Diedenhoven's survey planetary nebulae were
classified as class A sources. Tokunaga et al. (1991) used similar
criteria to categorize 3.3 $\mu$m emission, with their class 1
equivalent to the van Diedenhoven's class A$_{3.3}$. The results of this
classification are listed in Table 2. All sources with PAH
emission are of class A$_{3.3}$, as is consistent with both van
Diedenhoven and Tokunaga. The presence of strong atmospheric water
lines makes analysis of PAH emission line shape more difficult,
but all detected PAH features have FWHM and central emission wavelengths consistent with class A$_{3.3}$ emission, and appear to be largely symmetric. Other surveys of 3.3 $\mu$m emission in planetary nebulae
(Roche et al. 1996) were at lower resolution than the R=1700
spectra in this study. Our resolution allows us to fit and remove
the Pfund delta hydrogen emission line, which could otherwise
affect the fitted gaussian curve and apparent line shape.

The 3.3 $\mu$m feature shows relatively little variation even in
very different emission sources (van Diedenhoven 2004, Cohen \& Barlow 2005,
Tokunaga 1991), while longer wavelength PAH emission shows greater
variation in central wavelength and FWHM (Peeters et al. 2002). In
longer wavelength studies planetary nebulae are classified as
class B$_{6-9}$ objects, while some very young post-AGB stars exhibit
class C$_{6-9}$ emission. This apparent evolution in PAH emission during
the AGB to planetary nebula transition is the primary motivation
for expanding the sample of evolved stars investigated at high
resolution in all PAH bands.

In the current survey we did not sample multiple portions of the
target objects, instead sampling a 4-5 arcsec band across the
extent of the nebula. If there are more subtle changes in 3.3
$\mu$m emission due to nebular position, they would not be
detected by this approach.

\begin{figure}
\epsscale{1.1}
\plotone{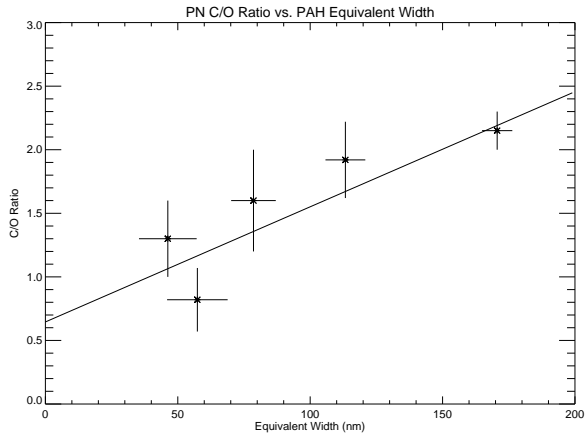}
\caption{PN C/O Ratio versus measured PAH EW. See Table 2 for data and references. The plotted line yields a threshold value of C$/$O=0.65 $\pm$ 0.28}.\label{fig4}
\end{figure}

\section{Conclusions}

With a resolution $\sim$1,700 we have carried out a preliminary
survey of 3.3 $\mu$m PAH emission using FLITECAM, an instrument
developed at UCLA, on the Lick Observatory 3-m telescope. This
survey is a pilot project for a deeper program to be carried out
using the same instrument on board the Stratospheric Observatory
for Infrared Astronomy (SOFIA).

(1)Out of 20 objects observed, 11 showed detectable 3.3$\mu$m PAH
emission.

(2)We found the 3.3$\mu$m feature to have a FWHM and central
wavelength consistent with that found by Tokunaga et al. (1991),
Roche et al. (1996) and van Diedenhoven et al. (2004).

(3)A correlation was found between C$/$O ratio and PAH emission
strength, consistent with that reported by Roche et al. (1996) and
Cohen \& Barlow (2005). We calculate a threshold C/O ratio of 0.65 $\pm$ 0.28.

(4)We classified the 11 objects showing PAH emission according to
the system suggested by van Diedenhoven.

Our high-resolution survey of 3.3$\mu$m emission has expanded the
sample of PNs with characterized PAH emission at this wavelength.
This step is important for evolutionary studies because shifts are
seen in longer wavelength PAH emission between PNs and their
evolutionary pre-cursors. It also serves as a starting point for
detailed studies of spatial variations of PAH emission in
individual planetary nebulae, as well as studies of PAH emission
in young AGB stars.

Although the sample of objects presented is relatively small,
these high-resolution 3.3$\mu$m PAH spectra provide a clear
indication of the importance of airborne astronomy to this field.
With SOFIA, the thermal background will be reduced by more than an
order of magnitude and none of the water vapor lines will
saturate. Higher signal-to-noise ratios will be possible, and it
may soon be practical to detect subtle changes in line shape due
to PAH ionization. Studies of the 3.3 $\mu$m band in Orion have
shown a shift from class A$_{3.3}$  to class B$_{3.3}$  emission with position (van Diedenhoven et
al. 2004), as well as a shift from symmetric to asymmetric
3.3$\mu$m emission. Similar shifts may occur in planetary nebulae,
and would be observable with lower backgrounds and water vapor
contributions. Likewise, in extended objects the lower backgrounds
will make it possible to probe the shape of the PAH feature across
the spatial extent of the nebula.

\acknowledgements 
The authors wish to thank the staff of the Lick
Observatory for their outstanding support. We also thank Eric
Becklin (UCLA) for his support throughout the development of the
FLITECAM instrument. E.C.S. is supported by a NASA Graduate Student Research Program (GSRP) fellowship. The authors wish to thank the anonymous referee, Dr. Mike Jura and Dr. Mark Morris for their  helpful suggestions in improving this manuscript.


\begin{thebibliography}{}
 \bibitem[Allamandola et al.(1989)]{1989ApJS...71..733A} Allamandola, L.~J., 
Tielens, G.~G.~M., \& Barker, J.~R.\ 1989, \apjs, 71, 733
\bibitem[Cohen \& Barlow(2005)]{2005MNRAS.362.1199C} Cohen, M., \& Barlow, M.~J.\ 2005, \mnras, 362, 1199 
\bibitem[Cohen et al.(1989)]{1989ApJ...341..246C} Cohen, M., Tielens, A.~G.~G.~M., Bregman, J., Witteborn, F.~C., Rank, D.~M., Allamandola, L.~J., Wooden, D., \& Jourdain de Muizon, M.\ 1989, \apj, 341, 246
\bibitem[De Marco \& Barlow(2001)]{2001Ap&SS.275...53D} De Marco, O., \& Barlow, M.~J.\ 2001, \apss, 275, 53
\bibitem[Geballe et al.(1985)]{1985ApJ...292..500G} Geballe, T.~R., Lacy, J.~H., Persson, S.~E., McGregor, P.~J., \& Soifer, B.~T.\ 1985, \apj, 292, 500 
\bibitem[Gillett et al.(1975)]{1975ApJ...200..609G} Gillett, F.~C., Forrest, W.~J., Merrill, K.~M., Soifer, B.~T., \& Capps, R.~W.\ 1975, \apj, 200, 609 
\bibitem[Gillett et al.(1973)]{1973ApJ...183...87G} Gillett, F.~C., Forrest, W.~J., \& Merrill, K.~M.\ 1973, \apj, 183, 87   
\bibitem[Hony(2002)]{2002PhDT..........H} Hony, S.\ 2002, Ph.D.~Thesis
\bibitem[Jourdain de Muizon et al.(1990)]{1990A&AS...83..337J} Jourdain de Muizon, M., Cox, P., \& Lequeux, J.\ 1990, \aaps, 83, 337 
\bibitem[Justtanont et al.(1996)]{1996ApJ...456..337J} Justtanont, K., Skinner, C.~J., Tielens, A.~G.~G.~M., Meixner, M., \& Baas, F.\ 1996, \apj, 456, 33
\bibitem[Kholtygin(1998)]{1998A&A...329..691K} Kholtygin, A.~F.\ 1998, \aap, 329, 691 
\bibitem[Leger \& Puget(1984)]{1984A&A...137L...5L} Leger, A., \& Puget, J.~L.\ 1984, \aap, 137, L5
\bibitem[Liu et al.(2004)]{2004MNRAS.353.1251L} Liu, Y., Liu, X.-W., Barlow, M.~J., \& Luo, S.-G.\ 2004, \mnras, 353, 1251 
\bibitem[Liu et al.(2001)]{2001MNRAS.323..343L} Liu, X.-W., et al.\ 2001, 
\mnras, 323, 343 
\bibitem[Lord(1992)]{1992nstc.rept.....L} Lord, S.~D.\ 1992, NASA TM103957
\bibitem[Mainzer \& McLean(2003)]{2003ApJ...597..555M} Mainzer, A.~K., \& McLean, I.~S.\ 2003, \apj, 597, 555
\bibitem[Mainzer et al.(2004)]{2004ApJ...604..832M} Mainzer, A.~K., McLean, 
I.~S., Sievers, J.~L., \& Young, E.~T.\ 2004, \apj, 604, 832 
\bibitem[Matsuura et al.(2005)]{2005A&A...434..691M} Matsuura, M., et al.\ 2005, \aap, 434, 691
\bibitem[McLean et al.(2006)]{2006SPIE.6269E.168M} McLean, I.~S., Smith, E.~C., Aliado, T., Brims, G., Kress, E., Magnone, K., Milburn, J., Oldag, A., Silvers, T., \& Skulason, G. \ 2006, \procspie, 6269. 
\bibitem[Middlemass(1990)]{1990MNRAS.244..294M} Middlemass, D.\ 1990, \mnras, 244, 294
\bibitem[Peeters et al.(2002)]{2002A&A...390.1089P} Peeters, E., Hony, S., Van Kerckhoven, C., Tielens, A.~G.~G.~M., Allamandola, L.~J., Hudgins, D.~M., \& Bauschlicher, C.~W.\ 2002, \aap, 390, 1089
\bibitem[Pwa et al.(1986)]{1986A&A...164..184P} Pwa, T.~H., Pottasch, S.~R., \& Mo, J.~E.\ 1986, \aap, 164, 184
\bibitem[Rinehart et al.(2002)]{2002MNRAS.336...66R} Rinehart, S.~A., Houck, J.~R., Smith, J.~D., \& Wilson, J.~C.\ 2002, \mnras, 336, 66
\bibitem[Roche et al.(1996)]{1996MNRAS.280..924R} Roche, P.~F., Lucas, P.~W., Hoare, M.~G., Aitken, D.~K., \& Smith, C.~H.\ 1996, \mnras, 280, 924
\bibitem[Sellgren(1984)]{1984ApJ...277..623S} Sellgren, K.\ 1984, \apj, 277, 623 
\bibitem[Sloan et al.(2007)]{2007ApJ...664.1144S} Sloan, G.~C., et al.\ 2007, \apj, 664, 1144
\bibitem[Smith \& McLean(2006)]{2006SPIE.6269E..50S} Smith, E.~C., \& McLean, I.~S.\ 2006, \procspie, 6269, 50
\bibitem[Tokunaga et al.(1991)]{1991ApJ...380..452T} Tokunaga, A.~T., Sellgren, K., Smith, R.~G., Nagata, T., Sakata, A., \& Nakada, Y.\ 1991, \apj, 380, 452 
\bibitem[van Diedenhoven et al.(2004)]{2004ApJ...611..928V} van Diedenhoven, B., Peeters, E., Van Kerckhoven, C., Hony, S., Hudgins, D.~M., Allamandola, L.~J., \& Tielens, A.~G.~G.~M.\ 2004, \apj, 611, 928 

 \end{thebibliography}
\end{document}